# Identification and Characterization of a New Disruption Regime in ADITYA-U Tokamak

Soumitra Banerjee[1,2], Harshita Raj[1,2], Sk Injamul Hoque[1,2], Komal Yadav[1,2], Sharvil Patel[3], Ankit Kumar[1,2], Kaushlender Singh[1,2], Ashok Kumawat[1,2], Bharat Hegde[1,2], Subhojit Bose[1], Priyanka Verma[1], Kumudini Tahiliani[1], Asha Adhiya[1], Manoj Kumar[1], Rohit Kumar[1], Malay Bikash Chowdhuri[1], Nilam Ramaiya[1], Ananya Kundu[1,2], Suman Aich[1,2], Suman Dolui[1,2], K. A. Jadeja[1], K.M. Patel[1], Ankit Patel[1], Rakesh L. Tanna[1], And Joydeep Ghosh[1,2]

[1]Institute for Plasma Research, Gandhinagar, 382428, India
[2]Homi Bhabha National Institute, Training School Complex, Anushaktinagar, Mumbai 400094, India
[3]University of Virginia, Department of Physics, Charlottesville, USA

Email: soumitra.banerjee@ipr.res.in

## ABSTRACT

Disruptions continue to pose a significant challenge to the stable operation and future design of tokamak reactors. A comprehensive statistical investigation carried out on the ADITYA-U tokamak has led to the observation and characterization of a novel disruption regime. In contrast to the conventional Locked Mode Disruption (LMD), the newly identified disruption exhibits a distinctive two-phase evolution: an initial phase characterized by a steady rise in mode frequency with a nonlinearly saturated amplitude, followed by a sudden frequency collapse accompanied by a pronounced increase in amplitude. This behaviour signifies the onset of the precursor phase on a significantly shorter timescale. Clear empirical thresholds have been identified to distinguish this disruption type from conventional LMD events, including edge safety factor, current decay coefficient, current quench (CQ) time, and CQ rate. The newly identified disruption regime is predominantly governed by the (m/n = 2/1) drift-tearing mode (DTM), which, in contrast to typical disruptions in the ADITYA-U tokamak that involve both m/n = 2/1 and 3/1 modes, consistently manifests as the sole dominant instability. Initiated by core temperature hollowing, the growth of this mode is significantly enhanced by a synergistic interplay between a strongly localized pressure gradient and the pronounced steepening of the current density profile in the vicinity of the mode rational surface.

## 1. INTRODUCTION

Disruptions represent a significant threat to the operational stability of tokamaks as fusion reactors, primarily because they can cause severe thermal and mechanical damage to critical components such as limiters, first-wall structures, and other plasma-facing components (PFCs) [1]. It poses potentially disastrous consequences for future large-scale devices such as ITER, DEMO, and DTT [2]. The multifaceted implications of these events include material loss through melting and vaporization driven by extreme thermal loads, and structural damage or failure resulting from the high electromagnetic forces generated by rapid transient currents in adjacent conductive structures [3]. Over several decades, tokamak experiments have identified several distinct categories of disruptions [4], including displacement disruptions (VDEs) [5], density-limit disruptions [6], low edge safety factor disruptions, and beta-limit disruptions [7]. The disruption process is typically characterized by a sequence of phases: an initiating event, a precursor phase, a thermal quench, and finally, a current quench [3]. A broad spectrum of initiating events and precursor behaviours can lead to diverse current quench scenarios [8]. Various initiating mechanisms lead to the onset of the precursor phase, wherein the amplification of magnetohydrodynamic (MHD) instabilities, primarily driven by magnetic island dynamics, plays a crucial role in triggering disruptions.

Generally, the precursor phase involves edge cooling or central temperature hollowing, leading to pronounced redistribution of the plasma current density [9]. The steep gradient of current density inside the q=2 surface or near the mode rational surface destabilizes the mode in both cases, thereby enhancing the potential for disruption. The thermal quench phase is initiated by the exponential growth of instabilities developed during the precursor stage, resulting in full magnetic field stochastization. This transition abruptly enhances cross-field thermal transport with perpendicular heat diffusivity exceeding typical values by orders of magnitude ($\sim 10 - 1000 \ m^2/s$) [4][10]. The rapid thermal energy dissipation manifests as an abrupt plasma temperature collapse, clearly observable through the characteristic decay of soft X-ray emission signals. The disruption culminates in the current quench phase, initiated by a distinctive spike in plasma current ($I_p$). The plasma's inherent inductance resists abrupt current changes, inducing a compensatory loop voltage ($V_{loop}$) surge that temporarily sustains Ip. Subsequently, the current decays resistively on characteristic L/R timescales [11], typically ranging from milliseconds to tens of milliseconds in modern tokamaks.

The past few decades have witnessed a concerted global effort towards the development of robust disruption prediction models for tokamaks [12]. This endeavour predominantly employs machine learning methodologies, capitalizing on the extensive datasets acquired from diverse plasma diagnostics across numerous experimental campaigns conducted on various devices. Notably, predictive models have been proposed for ADITYA [13], EAST [14], JET [15], HL-2A [16], and several other tokamaks. Furthermore, the ITPA MHD topical group has compiled a multi-device database of disruption characteristics to identify commonalities in disruption and mitigation behaviors across a broad spectrum of tokamaks [17], aiming to enhance the fundamental understanding of tokamak disruption physics. The ADITYA-U tokamak exhibits a marked susceptibility to disruptions driven by the (m/n = 2,1) drift tearing mode (DTM) [18], which has been extensively studied across numerous operational campaigns [19]. The uncontrolled growth of the tearing mode is notably reduced when coupled with the drift mode [20]. Experimental analysis confirms that the dominant (m/n = 2,1) drift tearing mode (DTM) rotates in the electron diamagnetic direction, with the observed frequency spectrum closely confirming theoretical predictions derived from diamagnetic drift dynamics. The experimental validation of the mode's rotational behaviour yields critical insights into the underlying instability mechanisms responsible for disruptive transitions in ADITYA-U. ADITYA-U, being a medium-sized tokamak, serves as a suitable test bed for the investigation of plasma disruptions. Various studies, including those involving pellet injection, disruptions triggered by Supersonic Molecular Beam Injection (SMBI), have been successfully conducted on this device.

Several dedicated experiments have been conducted to investigate and control the rotation frequency of magnetohydrodynamic (MHD) modes. In devices such as ADITYA-U [21], HBT-EP [22], and J-TEXT [23] tokamaks, electrode biasing has been effectively employed to modify the mode rotation frequency. Additionally, many tokamaks and reversed-field pinch (RFP) devices are now equipped with single or multiple rows of non-axisymmetric resonant magnetic perturbation (RMP) coils [24], which have been shown to alter tearing mode dynamics, including rotation frequency. In the ADITYA-U tokamak, gas puffing has also been utilized to reduce the mode rotation frequency [19]. Notably, all these experiments are aimed at actively modifying tearing mode behaviour to improve plasma stability. Recent observations in the JET tokamak report disruptions during the current termination phase [9]. Prior to disruption, the (2/1) mode shows an initial increase in rotation frequency, followed by rapid mode locking. This behaviour is linked to temperature hollowing and current flattening in the core, which steepen the current density gradient near the rational surface—conditions that destabilize the mode and trigger disruption.

This paper presents the identification and detailed characterization of a new type of disruption, distinct from the conventional locked mode disruption (LMD) [25] in ADITYA-U. In typical LMD scenarios, the interaction between the magnetic island and the resistive wall leads to a cessation of island rotation, subsequent to a rapid increase in island width and culminating in disruption. In contrast, disruptions observed in ADITYA-U exhibit a unique temporal behavior, characterized by a gradual increase in the MHD mode frequency, while the mode amplitude reaches a nonlinear saturation level significantly prior to the onset of the disruption [26]. This sequence, characterized by rising frequency and early amplitude saturation, indicates an alternative instability mechanism that does not conform to the classical mode-locking behaviour. The characterization of this distinct disruption process provides valuable insight into precursor dynamics and offers potential avenues for improving disruption prediction and control in tokamak plasmas. A comprehensive statistical analysis distinguishes LMD from this new type of disruption, highlighting clear differences in their parametric regime and the severity of their disruptive impact. The study demonstrates that disruptions can be effectively characterized using the edge safety factor and the current decay coefficient. A clear distinction emerges in the current quench (CQ) time and CQ rate [8] between the two disruptive discharge types, with new type of disruption exhibiting a markedly higher CQ rate compared to LMD, thereby posing a greater threat than conventional locked mode events. A comprehensive analysis of pre-disruptive mode frequency and edge safety factor reveals a strong correlation, while the current decay coefficient also shows a pronounced dependence on the pre-disruptive frequency. Specifically, elevated pre-disruptive frequencies are linked to shorter CQ durations, underscoring the direct impact of mode dynamics on disruption severity. Further examination highlights that enhanced core radiation and the development of a hollow temperature profile are prominent in this new type of disruptions, leading to modifications in the current density distribution [9]. This restructuring is likely to destabilize the (m/n = 2/1) mode, ultimately precipitating the onset of disruption.

The paper is structured as follows: Section 2 outlines the experimental setup and analysis techniques, Section 3 focuses on the observation and characterization of disruptions, Section 4 examines the dynamics of MHD frequency and explores potential reasons for its increase, and Section 5 concludes with a discussion and summary.

## 2. EXPERIMENTAL SETUP

The experiments are carried out in the ADITYA-U tokamak [27], a medium-sized, ohmically heated device with an air-core transformer configuration. The ADITYA-U tokamak employs a conventional toroidal configuration with major radius R = 0.75 m and minor radius a = 0.25 m, corresponding to an aspect ratio of R/a = 3. A real-time feedback control system maintains plasma equilibrium by adjusting magnetic fields to stabilize its horizontal position, allowing sustained discharges [28]. The plasma is limited by a circular graphite limiter composed of toroidal belts and poloidal quarter rings. The device operates with the following plasma parameters: plasma current ($I_p$) of 80–200 kA, toroidal magnetic field ($B_t$) of 0.9–1.2 T, chord-averaged electron density ($n_e$) of (1–3) × $10^{19} m^{-3}$, and central electron temperature ($T_e$) of 250–300 eV. The base vacuum pressure is maintained at approximately 3 × $10^{-8}$ Torr, while hydrogen plasma is generated at an operational pressure between 1 and 2 × $10^{-4}$ Torr. The present study examines discharges characterized by an edge safety factor ($q_{edge}$) between 3 and 6, indicating that the q = 2 surface is located near the plasma boundary in some shots, while positioned further inward in others. Standard diagnostic systems have been employed to measure the plasma parameters presented in this study. The line-averaged density is measured using a microwave interferometer, while the central electron temperature is inferred from soft X-ray emission ratios employing the foil-filter technique [29]. An array of twelve surface barrier detectors (ORTEC; active area ~50 mm², thickness ~100 μm), integrated into an imaging system spanning multiple poloidal chords, records the soft X-ray emissions. Magnetohydrodynamic (MHD) activity is monitored through two arrays of 16 Mirnov coils each, poloidally uniform and installed at diametrically opposite toroidal locations (90° and 270°), enabling the detection of fluctuations. Each coil, consisting of 35 turns and a cross-sectional area of 2.9 cm², contributes to mode structure identification through singular value decomposition (SVD) analysis [30]. Additionally, external magnetic diagnostics provide measurements of loop voltage, plasma current, position, and stored energy [31], with data acquired at a sampling rate of 100 kHz. Toroidal rotation velocities are inferred from Doppler shift measurements of CVI impurity emissions at 529.05 nm, utilizing a 1.0-meter Czerny–Turner spectrometer (Princeton Instruments, AM 510 [32]. Complementary edge plasma diagnostics include a rake Langmuir probe array that provides floating potential measurements across radial positions at the edge, sampled at 1 MHz.

## 3. OBSERVATION AND CHARACTERIZATION OF DISRUPTION IN ADITYA-U TOKAMAK

### 3.1 Observation of Accelerated Mode Disruption: A Distinction from Locked Mode Behavior

In the ADITYA-U tokamak, disruptions occur frequently, often triggered intentionally. Analysis has identified a novel disruption type, distinctly different from the commonly observed LMD. This new type of disruption is termed Accelerated Mode Disruption (AMD). Figure 1 presents two representative discharges (AMD and LMD) from the ADITYA-U tokamak, showing the time evolution of plasma current ($I_p$), loop voltage ($V_{loop}$), the time derivative of the poloidal magnetic field, and soft X-ray emissions. In both cases, the characteristic sequence of events leading to disruption is identifiable. The onset is marked by precursor oscillations in the Mirnov signals and a distinct negative spike in the loop voltage.

The rapid collapse of the soft X-ray signal indicates the onset of the thermal quench, during which the plasma thermal energy is abruptly lost. This is followed by a sharp decline in plasma current, ultimately culminating in the complete disruption event. The temporal alignment of these signatures remains consistent across both types of discharges, underscoring the commonality of the disruption mechanism despite differences in their triggering conditions.

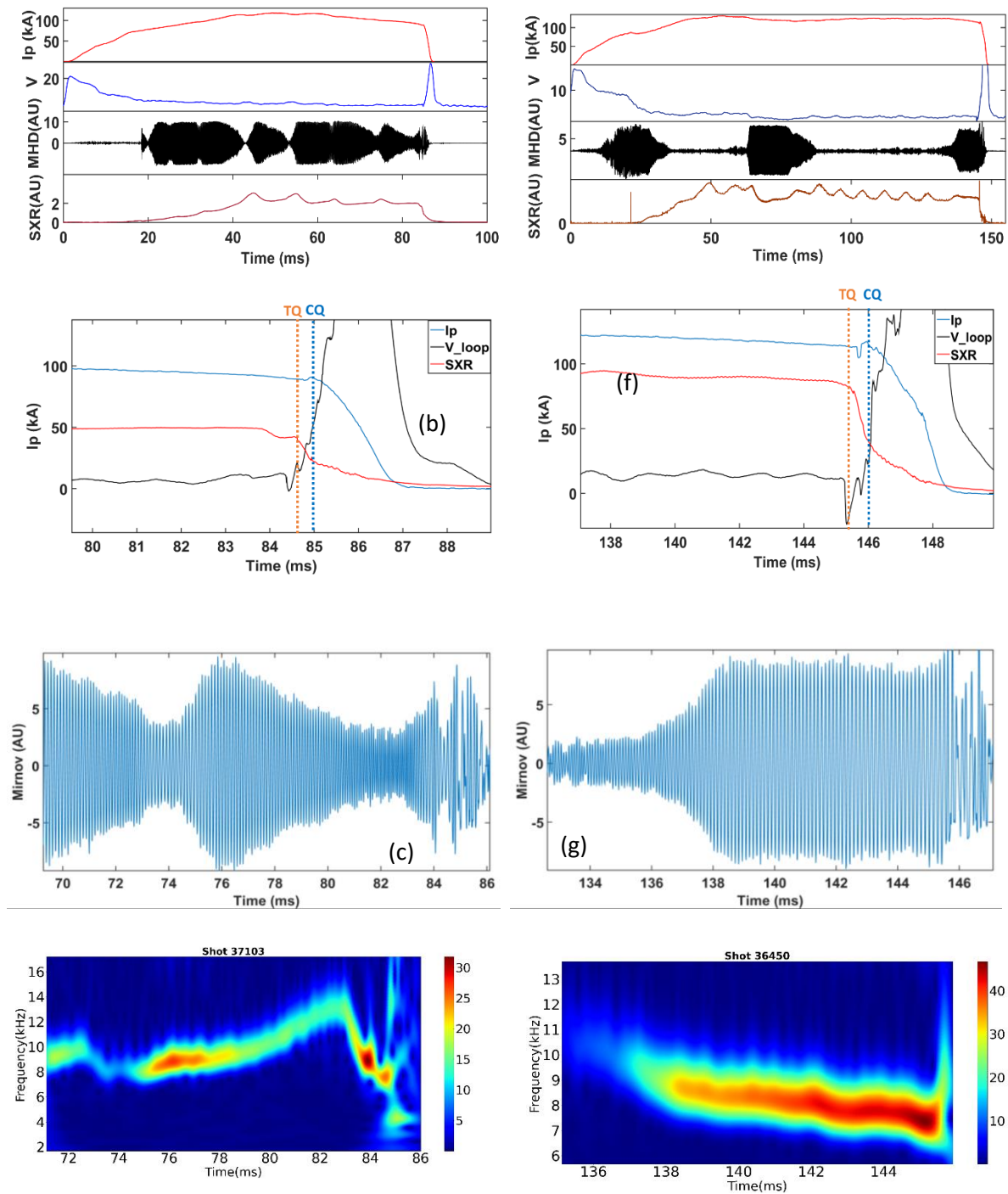

*Figure 1:* Temporal evolution of key plasma parameters for representative discharges of two disruption types in Aditya-U : Accelerated Mode Disruption (AMD) discharge #37103, showing plasma current (Ip), loop voltage (V), magnetic activity (MHD), and soft X-ray (SXR) signals(a). Disruption characteristics are inferred from variations in loop-voltage, SXR intensity, Ip decay(b), MHD amplitude(c) and frequency (d); Locked Mode Disruption (LMD) discharge #36450, with similar diagnostic traces shown in (e-h) for comparative analysis

In case of LMD, as seen in discharge #36450, the DTM frequency shows a gradual decay as the underlying instability grows. This continued slowdown eventually leads to mode locking, which serves as the immediate precursor to the disruption. In contrast, AMD exhibits distinctly different dynamics. Unlike locked mode disruptions (LMD), where the DTM frequency decays, AMD events are characterized by a monotonic frequency increase, indicating rotational acceleration. The instability nonlinearly saturates while the frequency continues to rise. This distinctive behaviour suggests that AMD is governed by fundamentally different mechanisms compared to conventional locked mode disruptions, with important implications for disruption prediction and mitigation strategies in tokamak operations. In AMD discharge #37103 (70–85 ms, Figure 1(d), the Mirnov signal shows a characteristic 50% increase in frequency, indicating sustained mode acceleration. In contrast, LMD discharge #36450 (132–150 ms, Figure 1(h)) exhibits a monotonic frequency decay in the Mirnov signal, reducing to 50% of its initial value is a signature of developing mode locking. While AMDs and LMDs display opposite trends in precursor frequency evolution (increasing versus decreasing), both follow an identical sequence leading to disruption.

## 3.2 Identification of the Parameter Regime Governing Accelerated Mode Disruption

Statistical analysis of disruptive discharges in the ADITYA-U tokamak reveals a distinct parameter space encompassing both LMD and AMD, enabling a clear distinction between these two disruption classes. The analysis identifies the edge safety factor ($q_{edge}$) which depends on the toroidal magnetic field ($B_t$), poloidal magnetic field ($B_p$) and so on, plasma current ($I_p$), along with the current decay coefficient as the dominant control parameter governing the disruption dynamics. The evolution of plasma current ($I_p$) for the representative AMD discharge (#37103) is depicted in Figure 2, providing insight into the current dynamics associated with this disruption type. The maximum plasma current ($I_{pmax}$) is identified during the flat-top phase, representing the peak attainable current level. The pre-disruptive current ($I_{pd}$) is determined at the onset of disruption, typically marked by a characteristic positive spike in Ip immediately preceding the current collapse.

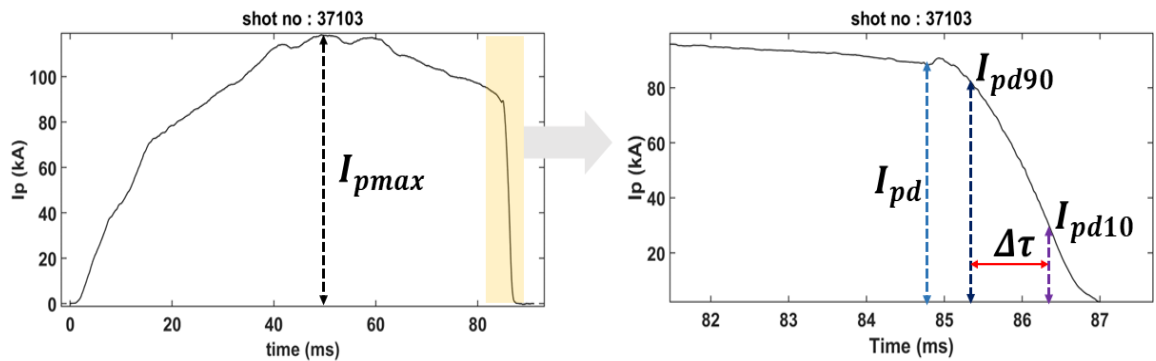

*Figure 2: Plasma current characterization for the disruptive discharges analyzed in this study*

To quantify the current quench (CQ) phase, $I_{pd90}$ and $I_{pd10}$ are defined as the plasma current levels corresponding to 90% and 10% of the pre-disruptive plasma current ($I_{pd}$) value. The duration of this CQ interval is influenced by the eddy currents induced in the conductive structures of the vacuum vessel during the current quench phase, which play a significant role in the dynamics of current decay. In ADITYA-U, the vessel is equipped with two toroidally separated electrical breaks, positioned 180 degrees apart, which introduce discontinuities in the vessel current paths. Consequently, measuring the CQ duration between $I_{pd90}$ and $I_{pd10}$ offers a robust and consistent metric for characterizing the current quench phase, as it is largely unaffected by the contributions from the vessel current.

In this analysis, the edge safety factor ($q_{edge}$) is evaluated at the time corresponding to the pre-disruptive plasma current ($I_{pd}$), while the normalized current decay coefficient (Z) is calculated in parallel. A comprehensive statistical analysis of 150 discharges was conducted, and the results were plotted, revealing two distinct clusters within the parameter space. When mapped against $q_{edge}$ at $I_{pd}$ and the normalized current decay coefficient, LMD and AMD cases exhibit separate and well-defined trends. This separation underscores the inherent physical differences between these two disruption mechanisms, highlighting their distinct characteristics.

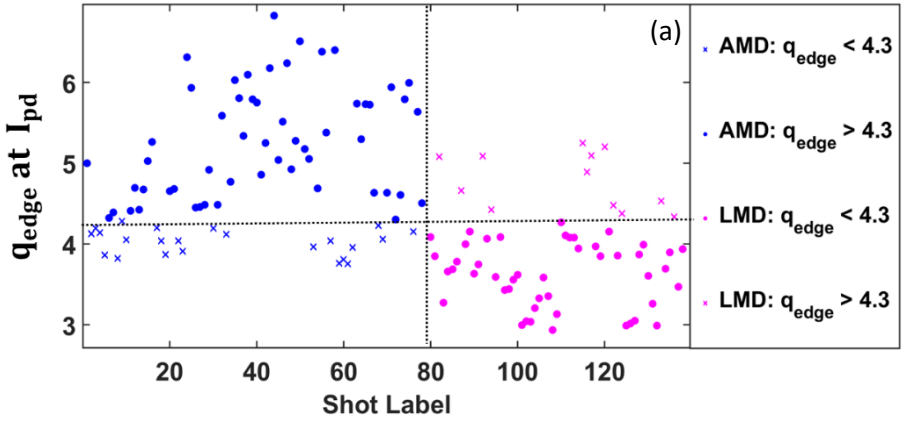

Fig 3 : **(a)** Edge safety factor ($q_{edge}$) evaluated at the disruption onset current (Ipd) for AMD and LMD cases, against shot labels **(b)** Statistical data of the current decay coefficient(Z) for AMD and LMD discharges with shot label

As illustrated in Fig. 3a, the statistical analysis demonstrates a strong correlation between the edge safety factor ($q_{edge}$) and the tendency toward AMDs or LMDs. Specifically, 73% of AMDs occur at $q_{edge} > 4.3$, while 80% of LMDs are observed when $q_{edge}$ is below 4.3. Discharges that lie within the transition regime exhibit a mix of behaviours, suggesting a more complex relationship that bridges the characteristics of both disruption types. These results highlight the significant role of the edge safety factor in distinguishing between AMD and LMD events, underscoring its importance as a key parameter in disruption dynamics.

Further analysis of the normalized plasma current decay coefficient, shown in Fig. 3b, reveals a strong dependence on disruption type. When the plasma current decreases by more than 16% from its maximum value, there is a 75% likelihood of an AMD occurring. On the other hand, when the current drop is less than 16%, the probability of an LMD occurring increases to 76%. This trend underscores the importance of the current decay coefficient in determining the nature of the disruption. A more pronounced reduction in current favors AMDs, while a smaller decay is more strongly associated with LMDs, suggesting that the magnitude of current decay is a critical factor influencing disruption mechanisms in the ADITYA-U tokamak.

## 3.3 Comparative Study of Current Quench Time and Rate in AMD and LMD Disruptions

A systematic evaluation of the current quench parameters, including the quench time ($\Delta\tau$) and quench rate, was conducted for both disruption classes. The CQ time ($\Delta\tau$) is defined as the time interval during which the plasma current decreases from 90% to 10% of its pre-disruption current ($I_{pd}$) value, i.e, between $I_{pd90}$ and $I_{pd10}$. The CQ rate, which quantifies the rapidity of the current decay, is computed as $(I_{pd90}-I_{pd10})/\Delta\tau$, representing the average rate of current reduction during the quench phase. It is well established that MHD instabilities and impurity influx significantly influence current quench dynamics by increasing plasma resistivity.

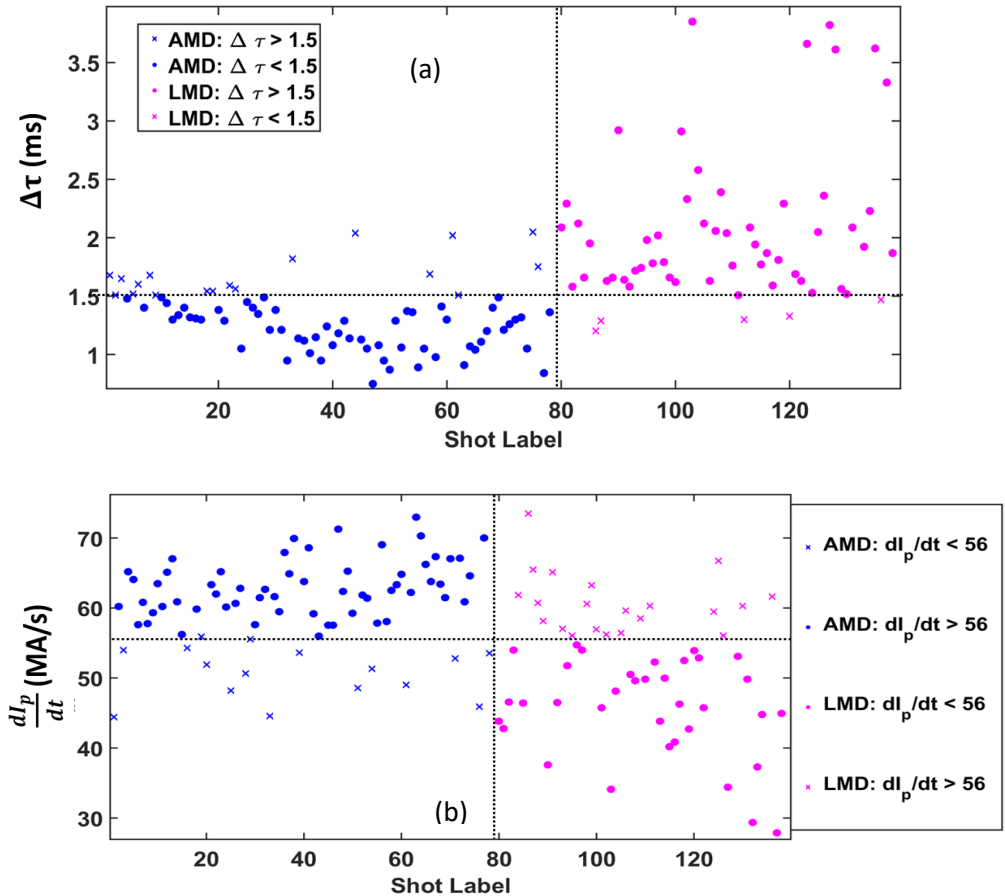

*Fig 4: (a) Current quench time for AMD & LMD discharges (b) Current quench rate for AMD & LMD discharges plotted against shot label*

A clear distinction in both quench rate and quench time has been observed between AMD and LMD events, indicating that different mechanisms govern the disruption process in each case. Figure 3 presents the statistical analysis of these parameters across 150 disruptive shots, clearly highlighting the differences between AMD and LMD events. The analysis reveals a clear contrast in current quench (CQ) times between AMD and LMD-type disruptions. AMD events exhibit notably shorter CQ times (<1.5 ms), whereas LMD events typically show longer durations (>1.5 ms). This distinction is further supported by the CQ rate, which exceeds 56 MA/s in AMD cases but remains below this threshold for LMD, as shown in Figure 3. Both excessively high and low current quench (CQ) rates pose significant challenges for tokamak operation. Elevated CQ rates, characteristic of AMD, impose significant thermal loads and mechanical stresses due to the rapid current collapse, rendering AMD particularly hazardous. Although a controlled plasma current ramp-down is ideal in theory, experimental observations reveal that during actual current decay, AMD frequently arises, precisely the condition where dangerously fast CQ dynamics (high rate, short duration) occur, posing serious operational risks that mitigation strategies seek to address.

## 4. DYNAMICS OF DRIFT-TEARING MODE FREQUENCY IN AMD: DEPENDENCE ON PARAMETRIC REGIMES AND ITS EFFECTS

The frequency of the drift-tearing mode (DTM) with mode numbers m/n = 2/1 shows a clear dependence on the edge safety factor ($q_{edge}$). An empirical correlation between $q_{edge}$ and the DTM frequency has been established specifically for AMD discharges, revealing a consistent and well-defined trend. This relationship is illustrated in the figure, which shows a linear dependence of the DTM frequency on $q_{edge}$.

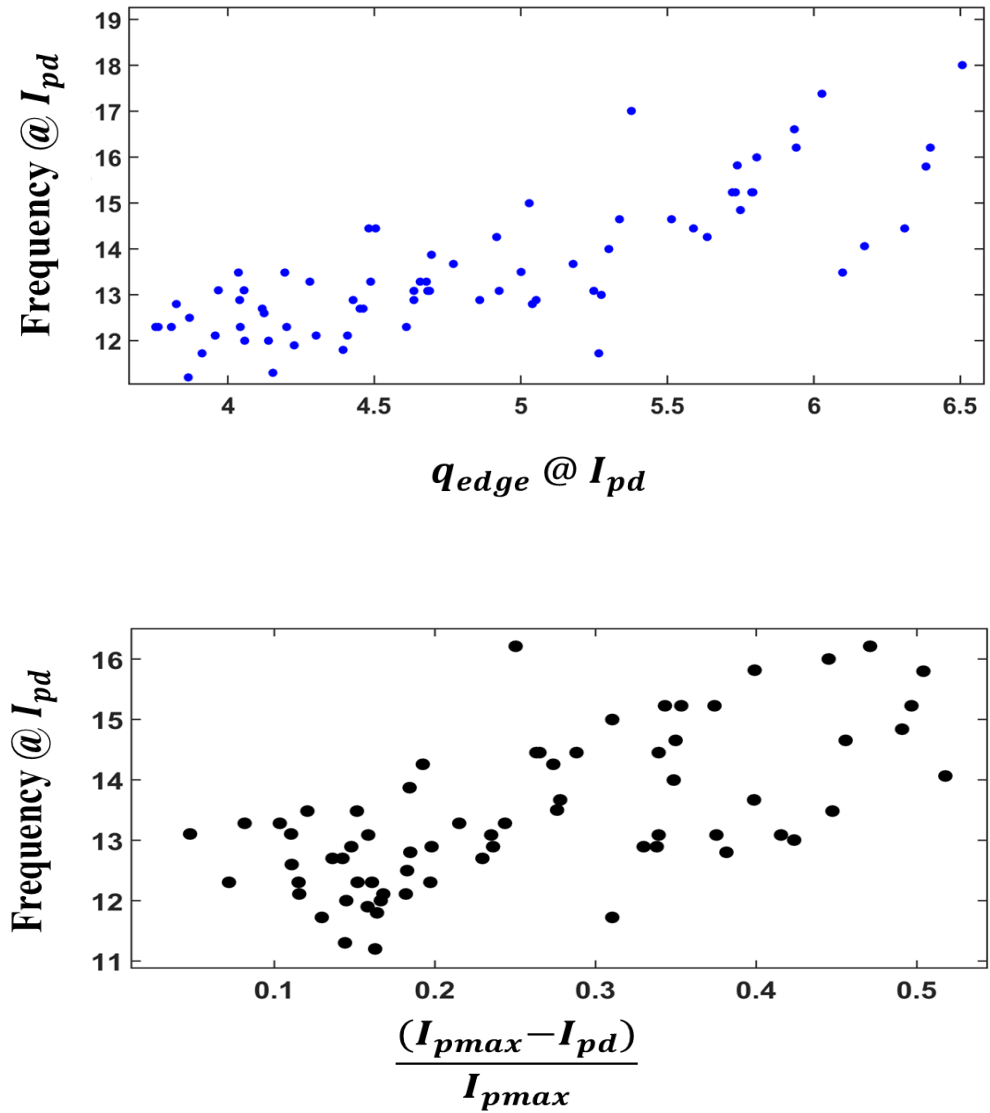

Fig 5: (a) DTM frequency as a function of edge safety factor ($q_{edge}$) at the pre-disruptive plasma current ($I_{pd}$) (b) DTM frequency plotted against the current decay coefficient

Furthermore, the second figure highlights a clear trend: a gradual decay of plasma current from its peak value corresponds to an increase in the DTM frequency. This pattern emphasizes the significant influence of current decay dynamics on the behaviour of the drift-tearing mode, indicating that a more gradual reduction in current favors the development of the mode at higher frequencies.

This correlation also suggests that current decay not only triggers MHD activity but also actively shapes the frequency characteristics of the evolving mode. During the current decay phase, the edge safety factor increases, leading to a flattening of the internal q-profile. Consequently, the q=2 surface shifts inward, and the rotation frequency of the m=2 mode increases.

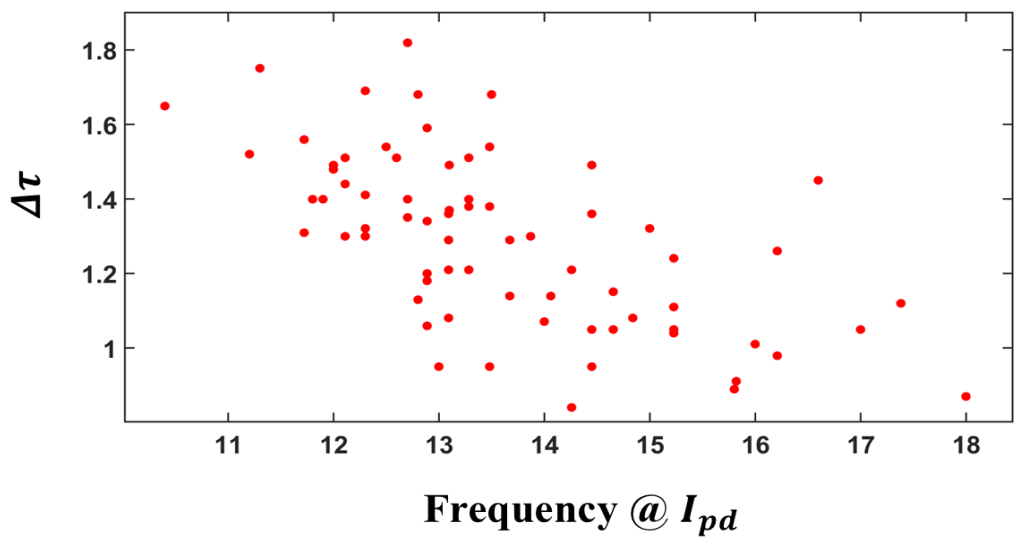

Fig 6: CQ time against DTM Frequency

A clear correlation between CQ time and DTM frequency at the time corresponding to $I_{pd}$, underscoring their mutual dependence, as shown in Figure 6. Specifically, discharges exhibiting higher DTM frequencies tend to show significantly shorter CQ times. This correlation suggests a potential coupling between the dynamics of the MHD mode and the rate of plasma energy dissipation during disruptions. It implies that as the DTM frequency increases, the plasma undergoes a more rapid current collapse, thereby accelerating the quench process. This observed interplay highlights the critical role of mode behavior in shaping the severity and timescale of disruptive events.

## 5. INFLUENCE OF KEY PARAMETERS ON DRIFT-TEARING MODE DYNAMICS

The magnetohydrodynamic (MHD) characteristics preceding both Accelerated Mode Disruptions (AMD) and Locked Mode Disruptions (LMD) have been systematically investigated using Mirnov coil measurements. This analysis incorporates multiple advanced analysis techniques to achieve a comprehensive understanding of mode dynamics. Initially, Fast Fourier Transform (FFT) analysis is applied to extract the dominant frequency components of the MHD activity, offering a preliminary spectral characterization of the modes. In addition, time-resolved frequency analysis is conducted using wavelet spectrograms, enabling the tracking of the temporal evolution of dominant mode frequencies and the identification of transient phenomena and frequency shifts[33]. Finally, Singular Value Decomposition (SVD) is employed to investigate the spatial structure of the modes, providing further insight into their underlying dynamics.

The analysis is based on an assumed current density profile of the form $J \sim \left(1 - \left(\frac{r}{a}\right)^2\right)^\alpha$, with α=3, The corresponding safety factor profile, q(r), is given by:

$$q(r) = q_{edge} \frac{\left(\frac{r}{a}\right)^2}{1-\left(1-\left(\frac{r}{a}\right)^2\right)^4} \quad \ldots\ldots\ldots\ldots (1)$$

Here $q_{edge} = \frac{2\pi a^2 B_t}{\mu_0 I_p R_0}$ denotes the edge safety factor. In the limit as r/a→0, the q(r) profile exhibits the relation $q_{edge} = (\alpha + 1).q(0)$, highlighting the scaling between the edge and core safety factors. The magnetic surface where the safety factor satisfies q=m/n (with m and n as integers) is referred to as the rational surface ($r_s$). The radial location of $r_s$ is primarily determined by the edge safety factor $q_{edge}$, exhibiting an inward shift with increasing $q_{edge}$ and an outward displacement as $q_{edge}$ decreases. This behavior highlights the sensitivity of the rational surface position to variations in the plasma q-profile. The width of the magnetic island associated with MHD modes at the rational surface radius $r_s$ can be estimated from the amplitude of the observed Mirnov coil signals[34], using the following relation:

$$W = 2r_c \sqrt{\left(\frac{2}{m}\right)\left(\frac{r_c}{r_s}\right)^2 \left(\frac{\widetilde{B_\theta}}{B_\theta}\right)} \quad \ldots\ldots\ldots\ldots (2)$$

Here, W represents the magnetic island width, m is the poloidal mode number, $r_s$ denotes the radial location of the resonant surface, and $r_c$ corresponds to the radial position of the Mirnov probe. The resonant surface radius $r_s$, for the (2,1) mode, is determined from the reconstructed safety factor (q) profile. For shot #37103, $r_s$, is estimated to be approximately 0.12 m. Analysis of the dominant MHD activity preceding the current quench (CQ) phase in AMD discharges reveals a strong (2,1) mode as the primary instability.

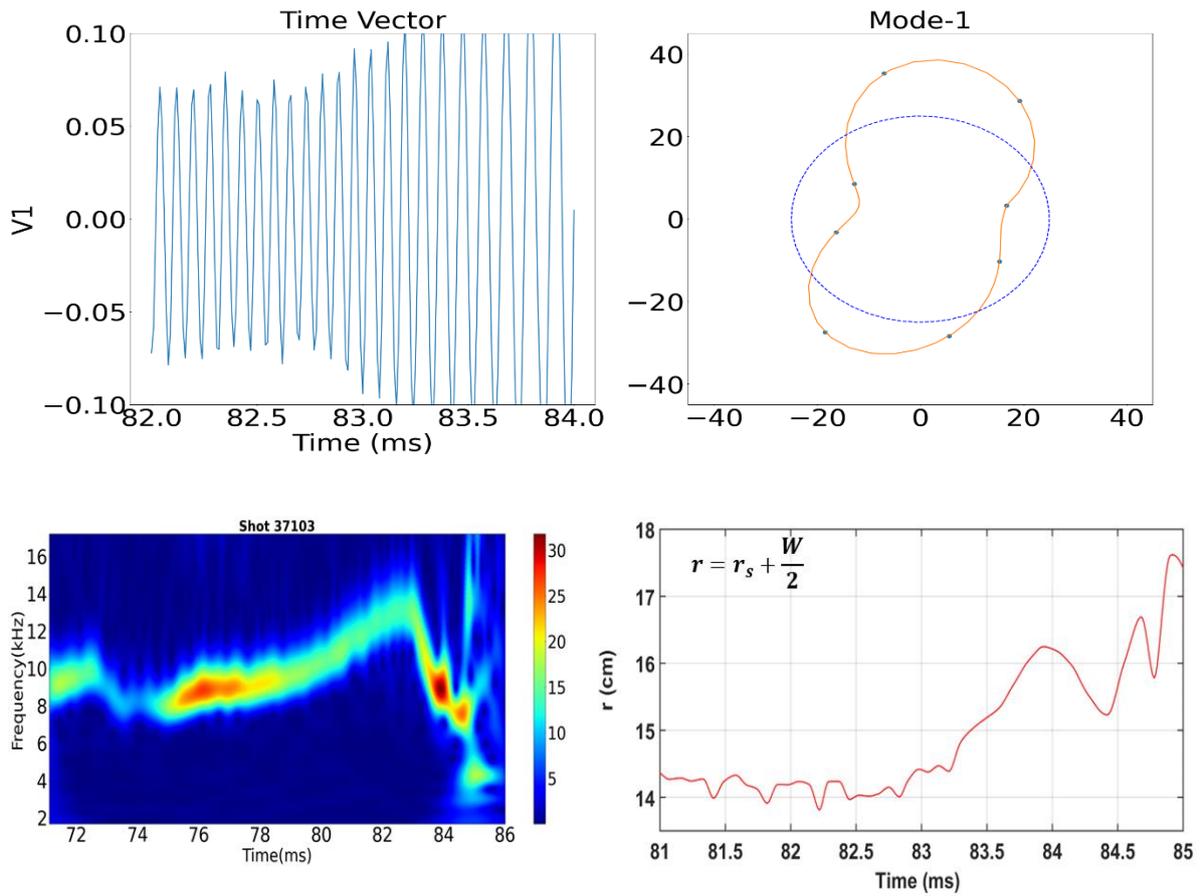

*Fig 7: (a) Time vector of Mirnov Signal (b) Spatial mode structure extracted using the Singular Value Decomposition (SVD) technique (c) Spectrogram showing the time evolution of mode frequency before and during the disruption phase (d) Estimated magnetic island width associated with the dominant MHD mode*
s

This mode exhibited pronounced activity during the pre-disruptive phase (~10 ms), characterized by a gradual increase in frequency. Prior to the onset of the precursor event, the width of the (2,1) magnetic island remained stable at approximately 4–5 cm and mode is located near 12.3 cm, until a sudden expansion occurred, coinciding with a sharp frequency drop. In shot #37103, the mode frequency begins to rise at t=68 ms, while the island width remains steady at 3–5 cm, with the frequency spanning 7–15 kHz up to t=83 ms. At this point, the frequency abruptly decreases from 13 kHz to 5 kHz within 1 ms, concurrent with a rapid expansion of the island width from 5 cm to 10 cm. These coupled phenomena ultimately precipitate the ensuing disruption event.

In general, the MHD frequency is defined as [35]:

$$f_{MHD} = \left[\frac{m}{2\pi r B_\varphi}\frac{\nabla p_e}{e\, n_e} + \frac{n v_\varphi}{2\pi R_0} + \frac{m v_\theta}{2\pi r}\right] \text{ at r = r'} \quad \ldots\ldots\ldots\ldots (3)$$

Here, r′ denotes the position of the rational surface, m and n represent the poloidal and toroidal mode numbers, respectively. $v_\varphi$ and $v_\theta$ are the toroidal and poloidal plasma flow velocities, $B_\varphi$ is the toroidal magnetic field, and $\nabla p_e$ corresponds to the electron pressure gradient. The plasma flow velocities and the electron pressure gradient strongly influence the frequency of the double tearing modes (DTMs). Specifically, the first term on the right-hand side of the governing equation (3) accounts for the diamagnetic contribution, whereas the second and third terms represent the effects of plasma flow. This relation indicates that an increase in the electron pressure gradient ($\nabla p_e$) directly contributes to a higher DTM frequency ($f_{MHD}$). Similarly, increases toroidal ($v_\varphi$) and poloidal ($v_\theta$) flow velocities act to further enhance the mode frequency.

For shot #37103, the toroidal velocity is plotted in the figure. Fig 8 reveals a strong correlation between the toroidal flow velocity and the mode frequency. When the MHD mode frequency is low, the toroidal velocity is approximately 5 km/s. As the disruption approaches, this velocity increases significantly to around 8 km/s, representing a 1.6-fold rise. Correspondingly, the mode frequency also increases, rising from 7 kHz to 15 kHz.

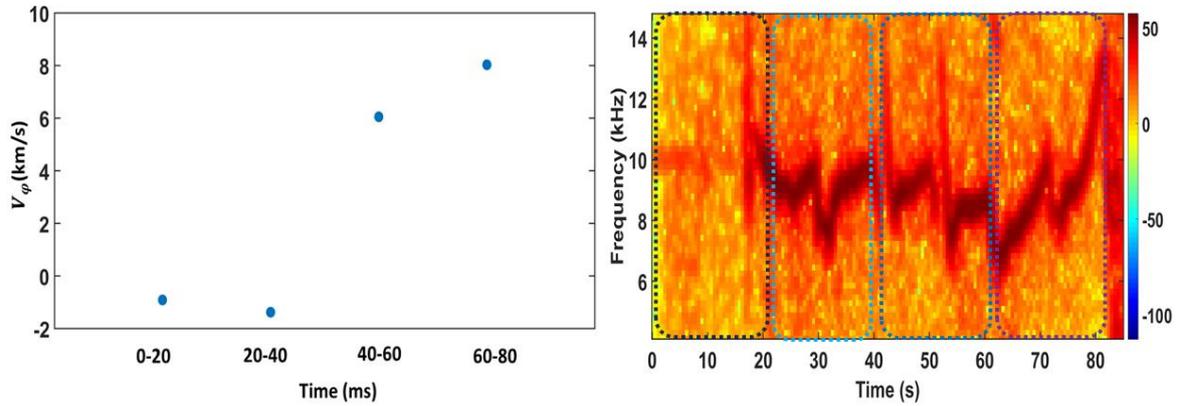

Fig 8 : Toroidal flow speed of plasma rotation

To further quantify the contribution of toroidal rotation to the total DTM frequency, the second term, representing the toroidal flow component, is calculated. This analysis allows for an assessment of the extent to which the background toroidal plasma rotation influences the overall observed frequency of the drift-tearing mode

$$\frac{nv_\varphi}{2\pi R_0} = \frac{1 \times 8000}{2 \times \pi \times 0.75} = 1.7 \, kHz$$

At 80 ms, the measured DTM frequency is approximately 10 kHz, while the toroidal flow component contributes around 1.7 kHz, accounting for roughly 17% of the total frequency.

To determine the poloidal flow velocity, the radial electric field ($E_r$) is required. Three reproducible discharges with identical plasma parameters were selected and analyzed to evaluate $E_r$. The average $E_r$ value obtained from these reference shots is used in the present analysis to ensure consistency. For these measurements, radial rake Langmuir probes designated LP-B and LP-C—were strategically positioned at radial locations of 23.9 cm and 24.7 cm, respectively, to facilitate accurate $E_r$ estimation. It is noteworthy that the radial electric field ($E_r$) undergoes a clear sign reversal that aligns with the onset of the DTM frequency increase and remains negative until the emergence of the precursor oscillations. Throughout this phase, the $E_r$ is consistently observed within the range of approximately −100 V/m to −300 V/m, indicating a sustained negative electric field that may play a role in influencing the mode dynamics during the pre-disruption phase.

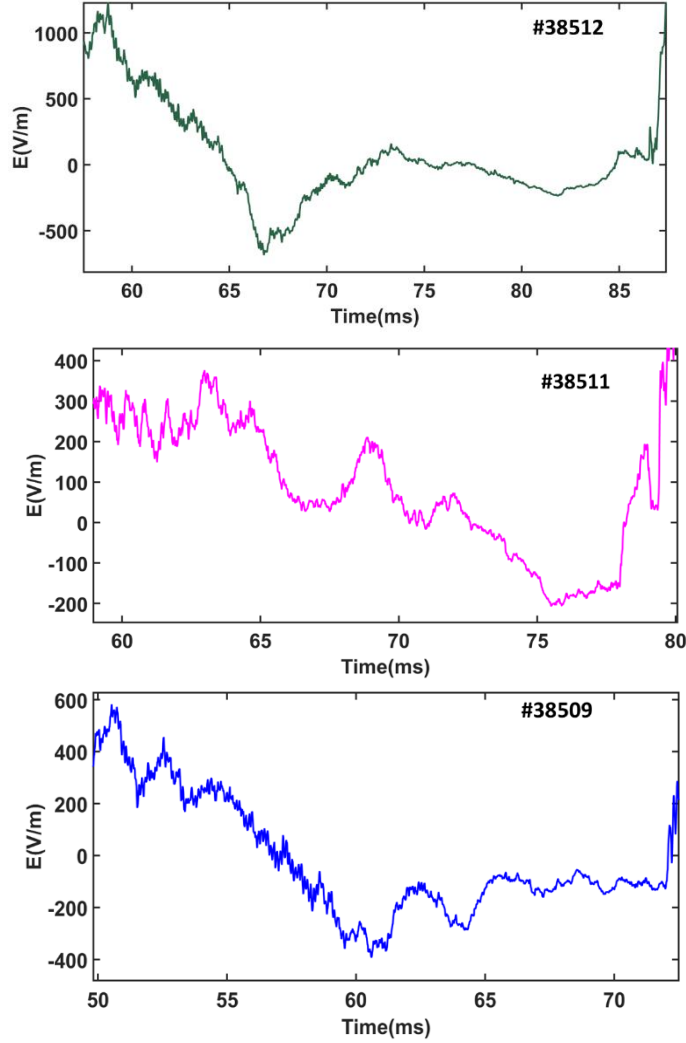

Fig 9 : Time evolution of radial E-field during frequency increase (AMD)

Subsequently, the third term, corresponding to the contribution of the poloidal flow to the total DTM frequency, is evaluated to quantify its role in the overall mode dynamics.

$$\frac{mv_\theta}{2\pi r} = \frac{m}{2\pi r}\left(\frac{E_r}{B_\varphi}\right) = \frac{2 \times 200}{2\pi \times 0.12 \times 1} = 0.5 \ kHz$$

Therefore, the poloidal flow component's contribution to the DTM frequency is comparatively small, representing only about 5% of the total frequency. With this in mind, by incorporating these values into the final equation, the term that contributes the maximum is

$$\frac{m}{2\pi r B_\varphi}\frac{\nabla p_e}{e\ n_e} = 7.8 \ kHz.$$

Approximately 22% of the increase in Drift-Tearing Mode (DTM) frequency can be attributed to the combined effects of toroidal and poloidal plasma flows, while the remaining contribution primarily arises from the diamagnetic term. This indicates that the pressure gradient near the mode rational surface plays a dominant role in driving the frequency increase observed during Accelerated Mode Disruptions (AMD)

# 6. DISCUSSION

The tearing mode is generally characterized by rapid growth; however, its growth rate significantly diminishes upon coupling with drift modes. This coupled instability, known as the drift-tearing mode (DTM), emerges as the dominant instability in ADITYA-U and plays a critical role in governing disruption dynamics in the device. To elucidate the underlying mechanisms of the observed AMD-type disruptions, it is crucial to analyze the dynamics of the (m/n=2/1) mode and identify the key factors influencing its behavior. A similar pattern has been reported in the JET tokamak[9], where tearing modes typically emerge during the plasma termination phase, demonstrating analogous dynamics to those observed in AMD-type disruptions. To differentiate between AMD and LMD-type disruptive discharges, a statistical study was conducted, revealing a cutoff in the edge safety factor and current decay coefficient. Furthermore, the impact of disruption is primarily determined by the CQ time and CQ rate, which exhibit distinct behaviors for the two discharge types. The analysis indicates that AMD is more hazardous than LMD, as it is associated with faster and more intense disruption dynamics.

The cause of LMD-type disruptions is well understood, whereas the mechanisms behind AMD-type disruptions differ significantly. To investigate these underlying causes, a comprehensive study has been conducted. As part of this study, the radiation emitted from the plasma is measured using a bolometer diagnostic system equipped with 16 lines of sight (chords), enabling the estimation of chord-averaged radiative losses across various vertical regions of the plasma. Figure 5 illustrates selected chords corresponding to both the plasma edge and core regions. The data reveal that, in the case of AMD, the radiation loss predominantly originates from the plasma core rather than the edge, coinciding with the onset of the frequency increase. Specifically, the chords positioned at −h8 cm (B4), −5 cm (B11), −1.5 cm (B3), and +8 cm (B12) are designated as core bolometer chords, as they primarily capture radiation from the plasma core. In contrast, the remaining chords shown in the plots, located beyond ±12 cm from the center, correspond to the edge region.

The line-averaged temperature has been estimated using the soft X-ray (SXR) signal, with 16 chords providing line-averaged temperature data. The SXR chords are positioned from −12.75 cm to +12.75 cm. Figure 6 illustrates the chord-averaged temperature data and their radial variation at two distinct times: at 66 ms, when the frequency begins to increase, and at 76 ms, when the frequency reaches near its peak.

In summary, the bolometer and chord-averaged temperature data offer valuable insights into the disruption dynamics, particularly in the case of AMD-type events. The bolometer data reveal that radiation increases primarily in the core region rather than at the edge, while the temperature profile exhibits a hollow structure, with the peak shifting from the core to a radial position between 7 and 9 cm. These findings highlight significant changes in plasma behavior as disruptions approach, providing a deeper understanding of the underlying mechanisms at play in AMD-type discharges.

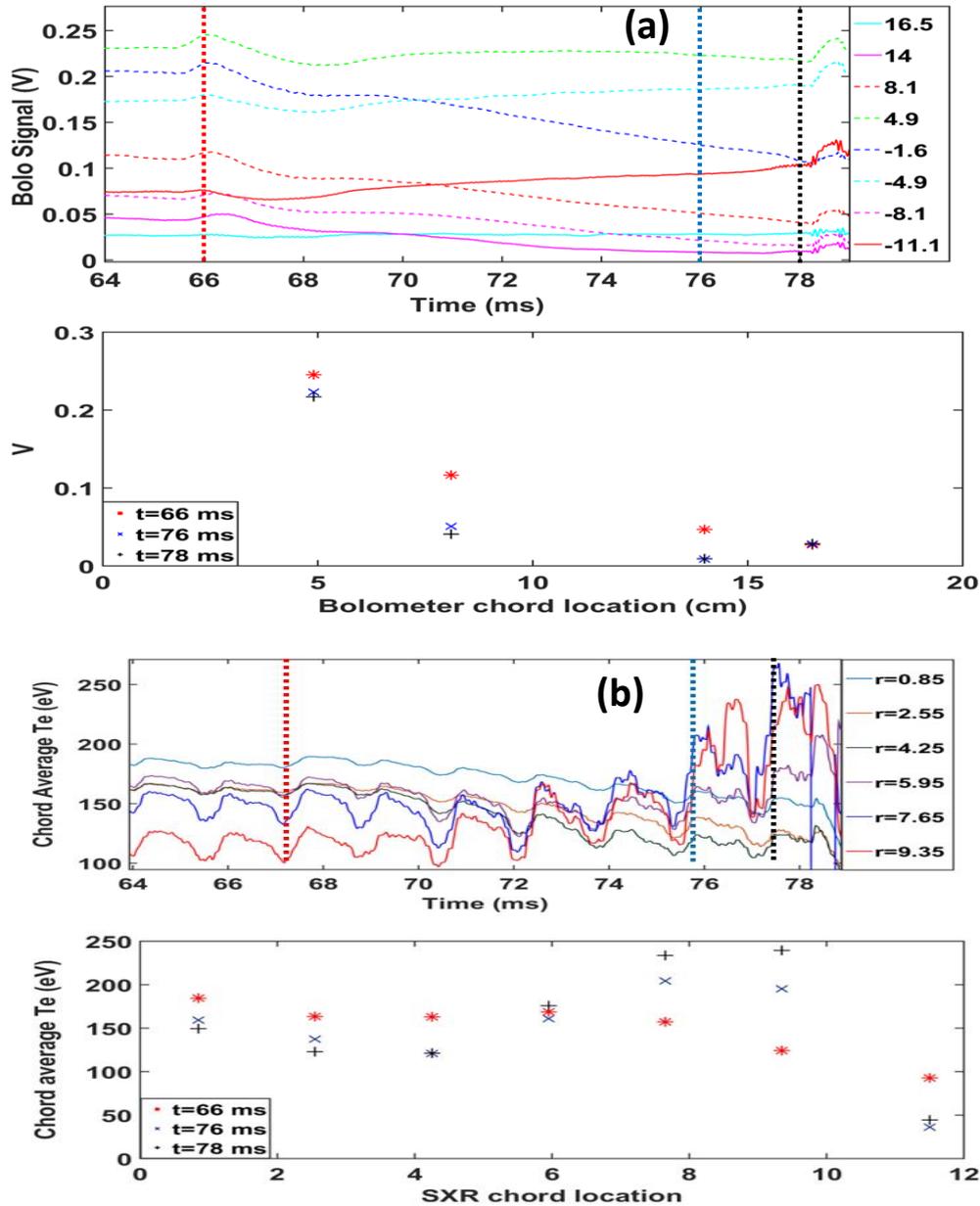

*Fig 10: (a)Bolometer radiation emission profile (b) Temperature evolution with time for AMD and chord average temperature with chord position*

As previously discussed, disruptions similar to AMD have also been reported in the JET tokamak during the plasma current termination phase. In these instances, the rise in core radiation has been identified as a key factor leading to plasma temperature hollowing. This hollowing effect alters the current density profile by steepening the gradient near the resonant surface of the (2/1) tearing mode, thereby enhancing its instability and facilitating the onset of disruption. These observations from JET closely align with the present findings, suggesting a common underlying mechanism governing AMD-type disruptions in ADITYA-U. The hollowing of the temperature profile can lead to a broadening of the current density profile, rendering it flat or slightly hollow near the magnetic axis, while simultaneously steepening the gradient near the mode's resonant surface. Consequently, the destabilization arises from the inward broadening of the current density profile, which enhances the drive for instability at the resonant surface. It is well established that the destabilization of classical tearing modes in tokamaks is primarily driven by the radial gradient of the toroidal current density, which represents the free energy source for the instability. In the ADITYA-U tokamak, operating under ohmic conditions without external heating, significant changes in the current profile occur during the current decay phase. Specifically, when the plasma current ($I_p$) falls by more than 16% from its maximum value ($I_{pmax}$), the increase in resistivity causes the current profile to evolve on a relatively short diffusion timescale, mirroring the changes in the electron temperature profile. In AMD-type disruptions, the associated mode is marked by a rising frequency, indicative of rational surfaces shifting inward due to modifications in the q-profile. As noted in reference, it is worth mentioning that all 2/1 modes destabilized as a consequence of temperature hollowing exhibit a rapid initial rotation, supporting the observed dynamics in ADITYA-U.

However, in the ADITYA-U tokamak, the observed instability is not a purely resistive tearing mode but rather a drift-tearing mode (DTM), where the classical tearing mode couples with the drift mode. As demonstrated in the previous section, the DTM frequency for shot #37103 is predominantly governed by the electron pressure gradient, highlighting the critical role of pressure-driven effects in the mode dynamics.

For the observed DTM frequency of 7.8 kHz, the relation, $\frac{m}{2\pi r B_\varphi} \frac{\nabla p_e}{e\, n_e} = 7.8\, kHz$, suggests that the required electron pressure gradient $\nabla p_e$ should be $2 \times 10^4\, Pa/m$. This implies that the combined contributions from the density gradient and the temperature gradient must satisfy the condition:

- Now, $\nabla p_e = \frac{11000 \times B_\varphi \times e \times n_e \times 2 \times r \times \pi}{2} = 2 \times 10^4\, Pa/m$
- $\nabla p_e = \nabla(n_e T_e) = \frac{2.6 \times 10^4}{1.6 \times 10^{-19}} = 1.625 \times 10^{23}$ Pa/m
- $n_e \nabla(T_e) + \mathbf{T_e} \nabla(n_e) = \mathbf{1.625 \times 10^{23}}$ **Pa/m**

The combined influence of temperature and density gradients produces a pressure gradient of $2 \times 10^4\, Pa/m$, an exceptionally high value under ADITYA-U tokamak conditions. Such a strong global pressure gradient is physically unsustainable, suggesting localized steepening near the mode rational surface. This sharp local pressure gradient significantly enhances the drive for the drift-tearing mode (DTM), sustaining the high mode frequency observed in the pre-disruption phase. These findings underscore the critical role of localized pressure steepening, along with the steep current density gradient near the rational surface, in destabilizing the mode and driving its growth.

# 7. SUMMARY

Disruption continues to pose a significant challenge to the stable tokamak operation. Through comprehensive Through a comprehensive statistical study, a novel disruption mode has been identified, termed Accelerated Mode Disruption (AMD) in ADITYA-U tokamak. A clear cutoff has been identified in both the edge safety factor and normalized current decay coefficient. Specifically, when the edge safety factor exceeds 4.3, the likelihood of an AMD-type disruption increases by 73%. Additionally, when the plasma current decay exceeds 16% of the maximum plasma current ($I_{pmax}$), the probability of an AMD-type disruption rises to 75%. The statistical analysis also demonstrates that AMD is more dangerous than LMD due to its significantly shorter current quench time and much higher current quench rate. The identification of AMD is characterized by increased core radiation emission, rather than from the plasma edge, accompanied by temperature hollowing. This temperature hollowing leads to a flattening of the current density profile and a steep gradient, along with steepening of the pressure gradient near the mode rational surface, providing free energy that destabilizes the mode and ultimately triggers the disruption. Based on these statistical findings and the identification of AMD, appropriate mitigation strategies should be implemented to address these high-risk disruptions.

As a prospective extension of this study, a comprehensive disruption mitigation strategy is planned for implementation in the ADITYA-U tokamak. This approach will be informed by critical thresholds, including the edge safety factor ($q_{edge}$), the current decay coefficient, and the amplitude of the Drift-Tearing Mode (DTM) frequency. The proposed mitigation framework will integrate both software-driven predictive algorithms and hardware-based actuators, enabling real-time identification and suppression of precursors to Accelerated Mode Disruptions (AMD), thereby improving the device's operational stability and enhancing safety by mitigating the risks associated with high current quench rates.

## REFERENCES

[1] T. Jordan and D. Schneider, "Effects of an electrically conducting first wall on the blanket loading during a Tokamak plasma disruption," *Fusion Engineering and Design*, vol. 31, no. 4, pp. 313–321, Aug. 1996, doi: 10.1016/0920-3796(96)00526-1.

[2] T. Craciunescu, A. Murari, on behalf of JET Contributors, and the EUROfusion Tokamak Exploitation Team, "Parsimonious statistical techniques for the detection of drifts toward dangerous operational conditions in tokamaks," *Plasma Phys. Control. Fusion*, vol. 66, no. 9, p. 095008, Sep. 2024, doi: 10.1088/1361-6587/ad670a.

[3] F. C. Schuller, "Disruptions in tokamaks," *Plasma Phys. Control. Fusion*, vol. 37, no. 11A, pp. A135–A162, Nov. 1995, doi: 10.1088/0741-3335/37/11A/009.

[4] E. Matveeva, J. Havlíček, O. Hronova, V. Weinzettl, and A. Havranek, "Disruptions and Plasma Current Asymmetries in Tokamak Plasmas".

[5] A. Hassanein, T. Sizyuk, and M. Ulrickson, "Vertical displacement events: A serious concern in future ITER operation," *Fusion Engineering and Design*, vol. 83, no. 7–9, pp. 1020–1024, Dec. 2008, doi: 10.1016/j.fusengdes.2008.05.032.

[6] M. Greenwald *et al.*, "A new look at density limits in tokamaks," *Nucl. Fusion*, vol. 28, no. 12, pp. 2199–2207, Dec. 1988, doi: 10.1088/0029-5515/28/12/009.

[7] A. D. Turnbull *et al.*, "Low-n ideal MHD stability of tokamaks: Current and beta limits," *Nucl. Fusion*, vol. 29, no. 4, pp. 629–639, Apr. 1989, doi: 10.1088/0029-5515/29/4/008.

[8] S. Purohit *et al.*, "Characterization of the plasma current quench during disruptions in ADITYA tokamak," *Nucl. Fusion*, vol. 60, no. 12, p. 126042, Dec. 2020, doi: 10.1088/1741-4326/abb79c.

[9] G. Pucella et al., "Onset of tearing modes in plasma termination on JET: the role of temperature hollowing and edge cooling," *Nucl. Fusion*, vol. 61, no. 4, p. 046020, Apr. 2021, doi: 10.1088/1741-4326/abe3c7.

[10] Z. Yang et al., "The study of heat flux for disruption on experimental advanced superconducting tokamak," *Physics of Plasmas*, vol. 23, no. 5, p. 052502, May 2016, doi: 10.1063/1.4948494.

[11] Y. Zhang et al., "Characteristics of current quenches during disruptions in the J-TEXT tokamak," *Phys. Scr.*, vol. 86, no. 2, p. 025501, Aug. 2012, doi: 10.1088/0031-8949/86/02/025501.

[12] E. Aymerich, G. Sias, S. Atzeni, F. Pisano, B. Cannas, and A. Fanni, "MHD spectrogram contribution to disruption prediction using Convolutional Neural Networks," *Fusion Engineering and Design*, vol. 204, p. 114472, Jul. 2024, doi: 10.1016/j.fusengdes.2024.114472.

[13] A. Agarwal et al., "Deep sequence to sequence learning-based prediction of major disruptions in ADITYA tokamak," *Plasma Phys. Control. Fusion*, vol. 63, no. 11, p. 115004, Nov. 2021, doi: 10.1088/1361-6587/ac234c.

[14] B. H. Guo et al., "Disruption prediction using a full convolutional neural network on EAST," *Plasma Phys. Control. Fusion*, vol. 63, no. 2, p. 025008, Feb. 2021, doi: 10.1088/1361-6587/abcbab.

[15] E. Aymerich, G. Sias, F. Pisano, B. Cannas, A. Fanni, and the-JET-Contributors, "CNN disruption predictor at JET: Early versus late data fusion approach," *Fusion Engineering and Design*, vol. 193, p. 113668, Aug. 2023, doi: 10.1016/j.fusengdes.2023.113668.

[16] Z. Yang, F. Xia, X. Song, Z. Gao, S. Wang, and Y. Dong, "In-depth research on the interpretable disruption predictor in HL-2A," *Nucl. Fusion*, vol. 61, no. 12, p. 126042, Dec. 2021, doi: 10.1088/1741-4326/ac31d8.

[17] N. W. Eidietis et al., "The ITPA disruption database," *Nucl. Fusion*, vol. 55, no. 6, p. 063030, Jun. 2015, doi: 10.1088/0029-5515/55/6/063030.

[18] D. Biskamp, "Drift-tearing modes in a tokamak plasma," *Nucl. Fusion*, vol. 18, no. 8, pp. 1059–1068, Aug. 1978, doi: 10.1088/0029-5515/18/8/003.

[19] H. Raj et al., "Effect of periodic gas-puffs on drift-tearing modes in ADITYA/ADITYA-U tokamak discharges," *Nucl. Fusion*, vol. 60, no. 3, p. 036012, Mar. 2020, doi: 10.1088/1741-4326/ab6810.

[20] R. Fitzpatrick and F. L. Waelbroeck, "Drift-tearing magnetic islands in tokamak plasmas," *Physics of Plasmas*, vol. 15, no. 1, p. 012502, Jan. 2008, doi: 10.1063/1.2829757.

[21] T. Macwan et al., "Controlling the rotation of drift tearing modes by biased electrode in ADITYA-U tokamak," *Physics of Plasmas*, vol. 28, no. 11, p. 112501, Nov. 2021, doi: 10.1063/5.0059410.

[22] J. W. Brooks, I. G. Stewart, M. D. Boyer, J. P. Levesque, M. E. Mauel, and G. A. Navratil, "Mode rotation control in a tokamak with a feedback-driven biased electrode," *Review of Scientific Instruments*, vol. 90, no. 2, p. 023503, Feb. 2019, doi: 10.1063/1.5062271.

[23] H. Liu et al., "Effect of electrode biasing on m/n = 2/1 tearing modes in J-TEXT experiments," *Nucl. Fusion*, vol. 57, no. 1, p. 016003, Jan. 2017, doi: 10.1088/0029-5515/57/1/016003.

[24] W. Suttrop et al., "In-vessel saddle coils for MHD control in ASDEX Upgrade," *Fusion Engineering and Design*, vol. 84, no. 2–6, pp. 290–294, Jun. 2009, doi: 10.1016/j.fusengdes.2008.12.044.

[25] M. F. F. Nave and J. A. Wesson, "Mode locking in tokamaks," *Nucl. Fusion*, vol. 30, no. 12, pp. 2575–2583, Dec. 1990, doi: 10.1088/0029-5515/30/12/011.

[26] J. Loizu and D. Bonfiglio, "Nonlinear saturation of resistive tearing modes in a cylindrical tokamak with and without solving the dynamics," *J. Plasma Phys.*, vol. 89, no. 5, p. 905890507, Oct. 2023, doi: 10.1017/S0022377823000934.

[27] R. L. Tanna *et al.*, "Overview of recent experimental results from the ADITYA-U tokamak," *Nucl. Fusion*, vol. 62, no. 4, p. 042017, Feb. 2022, doi: 10.1088/1741-4326/ac31db.

[28] P. Gautam *et al.*, "Improved Horizontal Plasma Position Control Using c-RIO-Based Real Time System in Aditya-U," *IEEE Trans. Plasma Sci.*, vol. 52, no. 9, pp. 3809–3813, Sep. 2024, doi: 10.1109/TPS.2024.3474713.

[29] J. Raval, "DEVELOPMENT OF MULTIPURPOSE SOFT X-RAY TOMOGRAPHY SYSTEM FOR ADITYA-U.".

[30] D. Raju, R. Jha, P. K. Kaw, S. K. Mattoo, Y. C. Saxena, and A. Team, "Mirnov coil data analysis for tokamak ADITYA," *Pramana - J Phys*, vol. 55, no. 5–6, pp. 727–732, Nov. 2000, doi: 10.1007/s12043-000-0039-8.

[31] S. Aich *et al.*, "Design and measurements of the diamagnetic loop in Aditya-U tokamak," *Radiation Effects and Defects in Solids*, pp. 1–12, Aug. 2024, doi: 10.1080/10420150.2024.2378424.

[32] A. Kumar *et al.*, "The effect of impurity seeding on edge toroidal rotation in the ADITYA-U tokamak," *Nucl. Fusion*, vol. 64, no. 8, p. 086019, Aug. 2024, doi: 10.1088/1741-4326/ad4c5a.

[33] J. Dhongde, S. Pradhan, and M. Bhandarkar, "MHD mode evolutions prior to minor and major disruptions in SST-1 plasma," *Fusion Engineering and Design*, vol. 114, pp. 6–12, Jan. 2017, doi: 10.1016/j.fusengdes.2016.10.015.

[34] A. D. Cheetham, S. M. Hamberger, H. Kuwahara, A. H. Morton, and D. Vender, "Pre-disruption MHD activity in the LT-4 tokamak," *Nucl. Fusion*, vol. 27, no. 5, pp. 843–847, May 1987, doi: 10.1088/0029-5515/27/5/013.

[35] X. D. Feng, G. Zhuang, Z. J. Yang, J. S. Xiao, J. Chen, and X. W. Hu, "Observation of the bifurcation of tearing modes due to supersonic gas injected into the J-TEXT plasmas," *Physics Letters A*, vol. 378, no. 16–17, pp. 1147–1152, Mar. 2014, doi: 10.1016/j.physleta.2014.02.017.